\documentclass[11pt]{article}
\usepackage{verbatim}
\usepackage{amsmath,amssymb}
\makeatletter \@addtoreset{equation}{section} \makeatother

\newcommand{\noi}{\vspace{12pt}\noindent}
\newcommand{\beq}{\begin{equation}}
\newcommand{\eeq}{\end{equation}}
\newcommand{\bea}{\begin{eqnarray}}
\newcommand{\eea}{\end{eqnarray}}

\newcommand{\e}[1]{{(\ref{#1})}}
\newcommand{\eq}[1]{{eq.\ (\ref{#1})}}
\newcommand{\es}[2]{{(\ref{#1}) and (\ref{#2})}}
\newcommand{\eqs}[2]{{eqs.\ (\ref{#1}) and (\ref{#2})}}
\newcommand{\Ref}[1]{{Ref.~\cite{#1}}}

\newcommand{\equi}[1]{\stackrel{{#1}}{=}}

\newcommand{\ie}{{${ i.e., \ }$}}
\newcommand{\eg}{{${ e.g., \ }$}}

\newcommand{\cf}{{cf.\ }}

\newcommand{\wrt}{{with respect to }}
\newcommand{\wrtt}{{with respect to the }}

\newcommand{\wthot}{{with the help of the }}

\renewcommand{\~}{ \ }
\renewcommand{\=}{ \ = \ }

\renewcommand{\Tilde}{\widetilde}
\renewcommand{\Bar}{\overline}
\newcommand{\eps}{\varepsilon^{}}
\newcommand{\p}{\!{}^{}}
\newcommand{\q}{{}^{}}

\newcommand{\So}{S_{\rm o}}

\newcommand{\brst}{{\bf \rm s}}
\newcommand{\tot}{{\rm tot}}

\newcommand{\sdet}{{\rm sdet}}
\newcommand{\rank}{{\rm rank}}

\newcommand{\gh}{{\rm gh}}

\newcommand{\Hf}{\frac{1}{2}}
\newcommand{\ih}{{\scriptstyle{\frac{i}{\hbar}}}}
\newcommand{\hi}{{\scriptstyle{\frac{\hbar}{i}}}}
\newcommand{\Ih}{\frac{i}{\hbar}}
\newcommand{\Hi}{\frac{\hbar}{i}}

\newcommand{\twostack}[2]{\begin{array}{c} \lower.8ex\hbox{${#1}$}
                     \cr \raise.8ex\hbox{${#2}$} \end{array}}
\newcommand{\deder}[1]{\frac{ 
 \stackrel{\raise.2ex\hbox{$\leftarrow$}}{\delta^{r}}   } 
 {   \delta {#1}}  }
\newcommand{\dedel}[1]{\frac{ 
 \stackrel{\lower.3ex \hbox{$\rightarrow$}}{\delta^{\ell}}   }
 {   \delta {#1}}  }
\newcommand{\papar}[1]{\frac{  
 \stackrel{\raise.2ex\hbox{$\leftarrow$}}{\partial^{r}}   } 
 {   \partial {#1}}  }
\newcommand{\papal}[1]{\frac{ 
 \stackrel{\lower.3ex \hbox{$\rightarrow$}}{\partial^{\ell}}   }
 {   \partial {#1}}  }
\newcommand{\rpa}[1]{{ 
 \stackrel{\raise.2ex\hbox{$\leftarrow$}}{\partial^{r}_{#1}}   }}
\newcommand{\lpa}[1]{{ 
 \stackrel{\lower.3ex\hbox{$\rightarrow$}}{\partial^{\ell}_{#1}}  }}

\begin{document}
\thispagestyle{empty}
\title{\Large{\bf External Sources in Field--Antifield Formalism}}
\author{{\sc Igor~A.~Batalin}$^{a}$ and {\sc Klaus~Bering}$^{b}$ \\~\\
$^{a}$I.E.~Tamm Theory Division\\
P.N.~Lebedev Physics Institute\\Russian Academy of Sciences\\
53 Leninsky Prospect\\Moscow 119991\\Russia\\~\\
$^{b}$Institute for Theoretical Physics \& Astrophysics\\
Masaryk University\\Kotl\'a\v{r}sk\'a 2\\CZ--611 37 Brno\\Czech Republic}
\maketitle
\vfill
\begin{abstract}
We introduce external sources $J\p_{A}$ directly into the quantum master action
$W$ of the field--antifield formalism instead of the effective action. The 
external sources $J\p_{A}$ lead to a set of BRST-invariant functions $W^{A}$ 
that are in antisymplectic involution. As a byproduct, we encounter
quasi--groups with open gauge algebras. 
\end{abstract}
\vfill
\begin{quote}
PACS number(s): 03.70.+k; 11.10.-z; 11.10.Ef; 11.15.-q;  
\hrule width 5.cm \vskip 2.mm \noindent 
$^{a}${\small E--mail:~{\tt batalin@lpi.ru}} \hspace{10mm}
$^{b}${\small E--mail:~{\tt bering@physics.muni.cz}} \\
\end{quote}

\newpage

\section{Introduction}
\label{secintro}

\noi
Historically, several authors have worked on a formalism with external sources, 
\eg in Yang-Mills theories \cite{bf75}, or \eg for the effective action 
$\Gamma$ \cite{vilk84a,vilk84b,dewitt87,bb12}. Here we shall not consider the
effective action, but rather work directly in terms of the quantum master 
action $W^{0}$.

\noi
At first sight, it seems tempting to try to introduce external sources 
$J\p_{\alpha}$ in the field--antifield formalism in a naive manner by simply 
modifying the standard quantum master action
\beq
W^{0}\~\stackrel{?}{\longrightarrow}\~ W^{0}+J\p_{\alpha}\Phi^{\alpha}\~,
\label{naivewrong}
\eeq
where $\Phi^{\alpha}$ denote the fundamental field variables,
$\alpha \in \{ 1,2, \ldots, N\}$.
However, consistency (as we shall see in next Section) requires $\Phi^{\alpha}$
to be BRST--invariant. This is in general not the case, so a more 
sophisticated approach is clearly needed.

\section{BRST--Invariant $W^{A}$ Functions}
\label{secw}

\noi
In this paper, we suggest to use BRST--invariant functions
$W^{A}\!=\!W^{A}(\Gamma;\hbar)$ to multiply the external sources $J\p_{A}$:
\beq
W^{J} \= W^{0} + J\p_{A}W^{A}\~.  \label{wj}
\eeq
The actions $W^{J}$ and $W^{0}$ denote the quantum master action with and 
without external sources $J\p_{A}$, respectively.
The index $A \in \{ 1,2, \ldots, 2N\}$ runs over twice as many values
as the index $\alpha \in \{ 1,2, \ldots, N\}$ to reflect the full
antisymplectic phase space
$\Gamma^{A}\!=\!\{\Phi^{\alpha};\Phi^{*}_{\alpha}\}$.
The Grassmann parity and ghost number are
\beq
\eps_{A} \~:=\~\eps(\Gamma^{A})\=\eps(W^{A})\=\eps(J\p_{A})\~,
\eeq
\beq
\gh_{A} \~:=\~\gh(\Gamma^{A})\=\gh(W^{A})\=-\gh(J\p_{A})\~,
\eeq
\beq
\eps(W^{J})\=0\~,\qquad \gh(W^{J})\=0\~.
\eeq
{\sc Remark:}
The action $W^{J}$ could more generally be a power series expansion in the 
sources $J\p_{A}$, but we shall for simplicity assume in this paper that
$W^{J}$ only depends affinely on the $J\p_{A}$ sources, as indicated in
\eq{wj}. (An {\em affine} function is a function with first--order terms and
zero-order terms.)

\noi
{\sc General remarks about Notation:} 
The superscript ``$0$'' on a quantity means the source--free limit $J\!=\!0$ of
that quantity. {}For example, $W^{0}=\left.W^{J}\right|_{J=0}$. 

\section{$\Delta$ operator and Antibracket $(\cdot,\cdot)$}
\label{secdeltaop}

\noi
To set up the field-antifield formalism \cite{bv81,bv83,bv84} one needs 
the $\Delta$ operator
\beq
\Delta\=\Delta\q_{\rho}+\nu\q_{\rho}\~,
\qquad \Delta^{2}\=0~,\qquad \eps(\Delta)\=1\~, 
\eeq
where  
\beq
\Delta\q_{\rho}\~:=\~
\frac{(-1)^{\eps_{A}}}{2\rho}\papal{\Gamma^{A}}\rho E^{AB}\papal{\Gamma^{B}}  
\eeq
is the odd Laplacian, and $\nu\q_{\rho}$ is an Grassmann--odd scalar function, 
see Refs.\ \cite{b06,b07,bb07,bb08,bb09} for details.

\noi
The antibracket is given as
\beq
(f,g)\~:=\~(-1)^{\eps_{f}}[[\stackrel{\rightarrow}{\Delta},f],g]1
\=(f\papar{\Gamma^{A}})E^{AB}(\papal{\Gamma^{B}}g)
\=-(-1)^{(\eps_{f}+1)(\eps_{g}+1)}(f\leftrightarrow g)\~.\label{antibracket01}
\eeq

\section{Quantum Master Equation}
\label{secqme}

\noi
The quantum master equation with external sources
\beq
\Delta e^{\ih W^{J}} \= 0\qquad \Leftrightarrow\qquad \Hf(W^{J},W^{J}) 
\= i\hbar\Delta\q_{\rho} W^{J}+\hbar^2\nu\q_{\rho}  
\label{qmej} 
\eeq
is equivalent to the following $J$-independent conditions 
\e{qme0}--\e{wainvol}.

\begin{enumerate}
\item
The standard quantum master equation:
\beq
\Delta e^{\ih W^{0}} \= 0\qquad \Leftrightarrow\qquad \Hf(W^{0},W^{0}) 
\= i\hbar\Delta\q_{\rho} W^{0}+\hbar^2\nu\q_{\rho}
\~.\label{qme0} 
\eeq
\item 
The functions $W^{A}\!=\!W^{A}(\Gamma;\hbar)$ are BRST--invariant,
\beq
\sigma\q_{W^{0}}(W^{A}) \= 0\~,\label{wabrstinv} 
\eeq
where $\sigma\q_{W^{0}}:=(W^{0},\~\cdot\~)+\hi \Delta\q_{\rho}$ is
the quantum BRST operator.
\item
The functions $W^{A}\!=\!W^{A}(\Gamma;\hbar)$ are\footnote{To see
\eq{wainvol}, differentiate the quantum master eq.\ \e{qmej} twice \wrtt
external sources $J\p_{A}$ and $J\p_{B}$ to get 
$(W^{A},W^{B})-(-1)^{(\eps_{A}+1)(\eps_{B}+1)}(A\leftrightarrow B)=0$.
Recalling the symmetry \e{antibracket01} of the antibracket then leads to
\eq{wainvol}.} 
mutually in involution \wrtt antibracket $(\cdot,\cdot)$,
\beq
(W^{A},W^{B}) \= 0\~.\label{wainvol}  
\eeq
\end{enumerate}

\noi
The third condition \e{wainvol} shows that the $2N$ function $W^{A}$ 
can carry at most $N$ independent functions, so in other words the
set $W^{A}$ will always be redundant. The redundant description is sometimes
necessary for relativistic quantum field theories to preserve symmetry, 
such as, \eg Lorentz symmetry, and locality. 

\noi
The second condition \e{wabrstinv} immediately illustrates 
that one cannot pick $W^{\alpha}\!=\!\Phi^{\alpha}$ and $W^{*}_{\alpha}\!=\!0$,
cf \eq{naivewrong}. This does not work because the fundamental field variables 
$\Phi^{\alpha}$ are in general not BRST invariant.

\section{Classical Master Equation}
\label{seccme}

\noi
Let us next consider the classical limit
\beq
W^{J}\=S^{J}+ {\cal O}(\hbar)\~,\qquad
W^{0}\=S^{0}+ {\cal O}(\hbar)\~,\qquad
W^{A}\=S^{A}+ {\cal O}(\hbar)\~,\label{wsexpand}
\eeq 
where
\beq
S^{J} \= S^{0}+ J\p_{A}S^{A}~.
\eeq
The classical master equation with external sources
\beq
(S^{J}, S^{J}) \= 0  \label{cmej}  
\eeq
is equivalent to the following $J$-independent conditions  
\beq
(S^{0}, S^{0}) \= 0 \~,\qquad
(S^{0}, S^{A}) \= 0 \~,\qquad
(S^{A}, S^{B}) \= 0 \~, \label{sainvol}  
\eeq
which, in turn, are the classical limit of the the conditions 
\e{qme0}--\e{wainvol}, respectively.

\section{Existence of $W^{J}$}
\label{secexistw}

\noi
Existence of the source--free classical master action $S^{0}$ for reducible 
theories was proven in \Ref{bv85} and further elaborated in \Ref{bb10}. The
presence of external sources $J\p_{A}$  does not change the proof in other
respect that pertinent quantities now depend on the external sources
$J\p_{A}$. A sufficient condition for the existence of the quantum master
action $W^{J}$ is that the cohomology of the classical BRST operator
$\brst^{J}$ vanishes in the sector with ghost number equal to $1$.

\section{Irreducible Theories}
\label{secirreduc}

\noi
We shall only consider the irreducible case from now on. According to
Theorem 3.4 of \Ref{bb10}, to prove the existence of the external source
formalism at the classical level, it remains to prove the existence of a
$J$-dependent acyclic, nilpotent Koszul--Tate operator $\brst_{-1}^{J}$. 
Here the nilpotency of $\brst_{-1}^{J}$ is just the $J$-dependent Noether
identities
\beq
(\So^{J}\papar{\varphi^{i}})\~R^{Ji}\q_{a}\=0\~,\label{noetherid01}
\eeq
where $\So^{J}=\So^{J}(\varphi)$ is the $J$-dependent action in the original
field sector, and $R^{Ji}\q_{a}=R^{Ji}\q_{a}(\varphi)$ are the $J$-dependent
gauge--generators. 

\noi
Thus we imagine that we are given a source--free theory that satisfies the 
Noether identity \e{noetherid01} for $J\p_{A}\!=\!0$, and we are seeking
solutions to these identities for non--vanishing external sources
$J\p_{A}\neq 0$.

\begin{table}[t]
\caption{Multiplicity, Grassmann parity and ghost number of the fundamental
variables $\Gamma^{A}$ and the BRST--invariant $W^{A}$ functions for irreducible
theories in the minimal sector.}

\label{multtable2}
\begin{center}
\begin{tabular}{|l|c||c|c||c|c|}  \hline
\multicolumn{2}{|c||}{}
&\multicolumn{2}{c||}{Fields $\Phi^{\alpha}$}&
\multicolumn{2}{c|}{Antifields $\Phi^{*}_{\alpha}$} \\ \hline
Variables&$\Gamma^{A}$&$\varphi^{i}$&$c^{a}$&
$\varphi^{*}_{i}$&$c^{*}_{a}$ \\ \hline
Multiplicity&$\rank(\Gamma^{A})\!=\!2N$&$n$&$m$&$n$&$m$ \\ \hline\hline
Grassmann Parity&$\eps_{A}$&$\eps_{i}$&$\eps_{a}\!+\!1$&$\eps_{i}\!+\!1\!$&
$\eps_{a}$ \\ \hline
Ghost Number& $\gh_{A}$&$0$&$1$&$-1$&$-2$ \\ \hline\hline
Rank of $W^{A}\!=\!S^{A}\!+\!{\cal O}(\hbar)$&$N\!=\!n\!+\!m$&
$n\!-\!m$&$m$&$m$&$0$\\ \hline
Classical BRST--Invariants&$S^{A}$&$S^{i}$&$S^{a}$&
$S^{*}_{i}$&$S^{*}_{a}\!\equiv\!0$ \\ \hline
Quantum BRST--Invariants&$W^{A}$&$W^{i}$&$W^{a}$&
$W^{*}_{i}$&$W^{*}_{a}\!\equiv\!0$ \\ \hline
\multicolumn{2}{|c||}{}
&\multicolumn{2}{c||}{$W^{\alpha}\!=\!S^{\alpha}\!+\!{\cal O}(\hbar)$}&
\multicolumn{2}{c|}{$W^{*}_{\alpha}\!=\!S^{*}_{\alpha}\!+\!{\cal O}(\hbar)$} 
\\ \hline
\end{tabular}
\end{center}
\end{table}

\section{Irreducible and Closed Theories}
\label{secirreducclosed}

\noi
In the irreducible and closed case, the proper solution can be taken on the 
form in the minimal sector
\beq
S^{J}\=\So^{J}+\varphi^{*}_{i}\~R^{Ji}\q_{a}\~c^{a}
+\Hf c^{*}_{c}\~U^{Jc}\q_{ab}\~c^{b}c^{a}(-1)^{\eps_{a}}\~,\label{closedaction}
\eeq
with $J$-dependent structure functions $U^{Jc}\q_{ab}=U^{Jc}\q_{ab}(\varphi)$. 
Besides the Noether identity \e{noetherid01}, the classical master equation 
\e{cmej} contains the gauge algebra relation
\beq
(R^{Ji}\q_{a}\papar{\varphi^{j}})\~R^{Jj}\q_{b}
-(-1)^{\eps_{a}\eps_{b}}(a \leftrightarrow b) 
\= R^{Ji}\q_{c}\~U^{Jc}\q_{ab}\~.\label{closedgaugealg}
\eeq
and a six--term Jacobi identity
\beq
\sum_{{\rm cycl.}\~a,b,c}(-1)^{\eps_{a}\eps_{c}}
\left(U^{Jd}\q_{ae}\~U^{Je}\q_{bc}
-(U^{Jd}\q_{ab}\papar{\varphi^{i}})\~R^{Ji}\q_{c}
\right)\= 0\~. \label{closedjacid}
\eeq

\section{Groupoid/Quasi--group}
\label{secgroupoid}

\noi 
The above set of eqs.\ \e{noetherid01}, \es{closedgaugealg}{closedjacid} has an
interpretation in terms of a (closed) groupoid/quasi--group \cite{b81}. The
fields $\varphi^{i}$ are coordinates on the quasi--group. We shall use the 
quasi--group construction to deduce BRST--invariants $S^{i}$ associated 
with the (transversal) original fields $\varphi^{i}$, \cf 
Section~\ref{secbrstinv}. (Differences in notation 
as compared with \Ref{bv81} and \Ref{b81} are for most parts obvious, except 
for the subtle fact that the structure functions
$U^{Jc}\q_{ab}=-t^{Jc}\q_{ab}$ have precisely the opposite sign there.) In
general, the quasi--group construction could in principle also works with
external sources $J\p_{A}$, as we will indicate in this
Section~\ref{secgroupoid}. However for the applications that we will present
in this paper in the next couple of Sections~\ref{secbrstinv}--\ref{secgi},
the external sources $J\p_{A}$ will actually not enter into the quasi--group
construction itself, but only have an organizing r\^{o}le (in the sense of
splitting the master equation in various sections).

\noi
Recall that the main idea of the quasi--group is to generalize Sophus Lie's 
original work for transformation groups, such that the composition law
$\Theta^{J}(\theta,\theta^{\prime};\varphi)$ for transformations (and hence
the structure ``constants'' $U^{Jc}\q_{ab}$) depend on the point $\varphi$. The
transformations (=arrows) are the (finite) gauge transformations
\beq
\varphi^{i}\~\longrightarrow\~
\Bar{\varphi}^{Ji}\=f^{Ji}(\varphi,\theta)\~,\label{phibar01}
\eeq
where $\theta^{a}$ are the gauge parameters. The composition law reads
\beq
f^{Ji}(f^{J}(\varphi,\theta),\theta^{\prime})
\=f^{Ji}(\varphi,\Theta^{J}(\theta,\theta^{\prime};\varphi))\~.
\label{storetheta01}
\eeq
The modified law of associativity reads
\beq
\Theta^{Ja}(\Theta^{J}(\theta,\theta^{\prime};\varphi),\theta^{\prime\prime};
\varphi)
\=\Theta^{Ja}(\theta,\Theta^{J}(\theta^{\prime},\theta^{\prime\prime};
f^{J}(\varphi,\theta));\varphi)\~. \label{ass01}
\eeq
The gauge transformation \e{phibar01} is assumed to have an inverse gauge 
transformation
\beq
\Bar{\varphi}^{i}\~\longrightarrow\~
\varphi^{Ji}\=((f^{J})^{-1})^{i}(\Bar{\varphi},\theta)\=
f^{Ji}(\Bar{\varphi},\vartheta(\theta;\Bar{\varphi}))
\~.\label{phibar02}
\eeq
Define
\bea
R^{Ji}\q_{a}(\varphi)
&:=&\left. (f^{Ji}(\varphi,\theta)
\papar{\theta^{a}})\q_{\varphi}\right|_{\theta=0}\~,\label{r01} \\
U^{Jc}\q_{ab}(\varphi)
&:=&\left.(\Theta^{Jc}(\theta,\theta^{\prime};\varphi)
\papar{\theta^{\prime a}}\papar{\theta^{b}})\q_{\varphi}
\right|_{\theta=0=\theta^{\prime}} 
-(-1)^{\eps_{a}\eps_{b}}(a \leftrightarrow b)\~,\label{u01}\\
\mu^{Ja}\q_{b}(\theta,\varphi)
&:=&\left.(\Theta^{Ja}(\theta,\theta^{\prime};\varphi)
\papar{\theta^{\prime b}})\q_{\varphi,\theta}\right|_{\theta^{\prime}=0}\~,
\qquad\qquad
\lambda^{J}\~:=\~(\mu^{J})^{-1}\~, \label{mulambda01}\\
\Tilde{\mu}^{Ja}\q_{b}(\theta,\varphi)
&:=&\left.(\Theta^{Ja}(\theta^{\prime},\theta;\varphi)
\papar{\theta^{\prime b}})\q_{\varphi,\theta}\right|_{\theta^{\prime}=0}\~,
\qquad\qquad
\Tilde{\lambda}^{J}\~:=\~(\Tilde{\mu}^{J})^{-1}\~,\label{mulambdatilde01}\\
\Sigma^{Ji}\q_{j}(\varphi,\theta)
&:=&(f^{Ji}(\varphi,\theta)\papar{\varphi^{j}})\q_{\theta}\~,\label{s01}\\
E^{J}(\varphi,\theta)&:=&\sdet(\Sigma^{J}(\varphi,\theta))\~
\frac{\sdet(\mu^{J}(\theta,\varphi))}{
\sdet(\Tilde{\mu}^{J}(\theta,\varphi))}
\~.\label{e01}
\eea
It is assumed that the matrices \e{mulambda01}, \es{mulambdatilde01}{s01} are
invertible. The Lie equation
\beq
(\Bar{\varphi}^{Ji}\papar{\theta^{b}})\q_{\varphi}
\~\equi{\e{bevislieeq01}}\~
R^{Ji}\q_{a}(\Bar{\varphi}^{J})\~\lambda^{Ja}\q_{b}(\theta,\varphi)\~,
\label{lieeq01}
\eeq
follows from
\bea
R^{Ji}\q_{b}(\Bar{\varphi}^{J})
&\equi{\e{r01}}&
\left. (f^{Ji}(\Bar{\varphi}^{J},\theta^{\prime})
\papar{\theta^{\prime b}})\q_{\varphi}\right|_{\theta^{\prime}=0}
\~\equi{\e{phibar01}+\e{storetheta01}}\~
\left.f^{Ji}(\varphi,\Theta^{J}(\theta,\theta^{\prime};\varphi))
\papar{\theta^{\prime b}}\right|_{\theta^{\prime}=0} \cr
&\equi{\e{mulambda01}}&
(\Bar{\varphi}^{Ji}\papar{\theta^{a}})\q_{\varphi}\~
\mu^{Ja}\q_{b}(\theta,\varphi)\~.\label{bevislieeq01}
\eea
The inverse Lie equation can be deduced as follows
\bea
-(\varphi^{i}\papar{\theta^{b}})\q_{\Bar{\varphi}^{J}}
&=&(\varphi^{i}\papar{\Bar{\varphi}^{Jj}})\q_{\theta}\~
(\Bar{\varphi}^{Jj}\papar{\theta^{b}})\q_{\varphi}
\~\equi{\e{s01}}\~
((\Sigma^{J})^{-1})^{i}\q_{j}(\varphi,\theta)\~
(\Bar{\varphi}^{Jj}\papar{\theta^{b}})\q_{\varphi} \cr
&\equi{\e{bevisinvlieeq01}}&
R^{Ji}\q_{a}(\varphi)\~\Tilde{\lambda}^{Ja}\q_{b}(\theta,\varphi)\~.
\label{invlieeq01}
\eea
In the last equality of \eq{invlieeq01} we used that 
\bea
\Sigma^{Ji}\q_{j}(\varphi,\theta)\~R^{Jj}\q_{b}(\varphi)
&\equi{\e{r01}+\e{s01}}&
(\Bar{\varphi}^{Ji}\papar{\varphi^{j}})\q_{\theta}
\left. (f^{Jj}(\varphi,\theta^{\prime})
\papar{\theta^{\prime b}})\q_{\varphi}\right|_{\theta^{\prime}=0}
\~\equi{\e{phibar01}}\~
\left.f^{Ji}(f^{J}(\varphi,\theta^{\prime}),\theta)
\papar{\theta^{\prime b}}\right|_{\theta^{\prime}=0} \cr
&\equi{\e{storetheta01}}&
\left.f^{Ji}(\varphi,\Theta^{J}(\theta^{\prime},\theta;\varphi))
\papar{\theta^{\prime b}}\right|_{\theta^{\prime}=0} \cr
&\equi{\e{phibar01}+\e{mulambdatilde01}}&
(\Bar{\varphi}^{Ji}\papar{\theta^{a}})\q_{\varphi}\~
\Tilde{\mu}^{Ja}\q_{b}(\theta,\varphi)\~.\label{bevisinvlieeq01}
\eea
Using similar arguments and, in particular, associativity \e{ass01}, it is 
possible to deduce the Maurer--Cartan equation and the inverse Maurer--Cartan
equation
\bea
(\lambda^{Ja}\q_{b}\papar{\theta^{c}})
-(-1)^{\eps_{b}\eps_{c}}(b \leftrightarrow c) 
&=&U^{Ja}\q_{de}(\Bar{\varphi}^{J})\~
\lambda^{Je}\q_{b}\~\lambda^{Jd}\q_{c}(-1)^{\eps_{b}\eps_{d}}\~, \\
(\Tilde{\lambda}^{Ja}\q_{b}\papar{\theta^{c}})
-(-1)^{\eps_{b}\eps_{c}}(b \leftrightarrow c) 
&=&-U^{Ja}\q_{de}(\varphi)\~\Tilde{\lambda}^{Je}\q_{b}\~
\Tilde{\lambda}^{Jd}\q_{c}(-1)^{\eps_{b}\eps_{d}}\~. 
\eea
It will become important when discussing quantum corrections in 
Section~\ref{secqcoor} that the $E^{J}$-function \e{e01} satisfies an initial 
value problem \cite{b81}
\beq
(\ln E^{J}(\varphi,\theta)\papar{\theta^{b}})\q_{\varphi}\=
A_{a}^{J}(f^{J}(\varphi,\theta))\~\lambda^{Ja}\q_{b}(\theta,\varphi)\~, \qquad 
E^{J}(\varphi,\theta\!=\!0)\=1\~, \label{epde01}
\eeq
which in turn satisfies pertinent consistency relations. Here we have defined 
the formal anomaly function
\beq
A_{a}^{J}\~:=\~(-1)^{\eps_{i}}(\papal{\varphi^{i}}R^{Ji}\q_{a})
+(-1)^{\eps_{b}}U^{Jb}\q_{ba}\~. \label{closedanomaly}
\eeq
Locally, \eq{epde01} leads to an integral representation
\beq
\ln E^{J}(\varphi,\theta)\= 
\int_{0}^{\theta} A_{a}^{J} (f^{J}(\theta^{\prime},\varphi))
\~\lambda^{Ja}\q_{b}(\theta^{\prime},\varphi)\~d\theta^{\prime b}\~, 
\label{eint01}
\eeq
where the integral \e{eint01} is independent of the integration contour.

\section{Construction of BRST--invariants $S^{i}$}
\label{secbrstinv}

\noi
Let the original action $\So^{0}$ be invariant under gauge transformations 
\e{phibar01}. We will for simplicity restrict our search to solutions 
$R^{Ji}\q_{a}\!=\!R^{0i}\q_{a}$ and $U^{Jc}\q_{ab}\!=\!U^{0c}\q_{ab}$ 
that are independent of the external sources $J\p_{A}$, so that the external 
sources only enter through the action 
\beq
S^{J}\=S^{0}+J\p_{A}\~S^{A}\=S^{0}+J\p_{i}\~S^{i}\~,\qquad 
S^{0}\=\So^{0}+\varphi^{*}_{i}\~R^{0i}\q_{a}\~c^{a}
+\Hf c^{*}_{c}\~U^{0c}\q_{ab}\~c^{b}c^{a}(-1)^{\eps_{a}}\~,
\label{minsjaction}
\eeq
in the original field sector, \ie via $J\p_{i}$. Here we will focus on 
constructing the BRST--invariants $S^{i}$ associated with the 
original fields $\varphi^{i}$ (or more precisely the transversal parts 
thereof). The idea is to gauge--fix the $m$ quasi--group gauge--parameters 
$\theta^{a}$ to be a function $\theta^{a}\!=\!\theta^{a}(\varphi)$ of 
$\varphi$ in precisely such a way that 
\beq
S^{i}\~:=\~\Bar{\varphi}^{0i}(\varphi,\theta(\varphi)) \label{siansatz01}
\eeq
become $n$ gauge--invariants, of which $n\!-\!m$ are independent. 
Total differentiation \wrt $\varphi^{j}$ yields
\beq
(S^{i}\papar{\varphi^{j}})
\~\equi{\e{siansatz01}}\~(\Bar{\varphi}^{0i}\papar{\varphi^{j}})\q_{\theta}
+(\Bar{\varphi}^{0i}\papar{\theta^{b}})\q_{\varphi}
(\theta^{b}\papar{\varphi^{j}})
\~\equi{\e{s01}}\~\Sigma^{0i}\q_{j}
+(\Bar{\varphi}^{0i}\papar{\theta^{b}})\q_{\varphi}
(\theta^{b}\papar{\varphi^{j}})\~. \label{sdiffwrtphi}
\eeq
Let $\chi^{a}\!=\!\chi^{a}(S^{i})$ be the $m$ independent gauge-fixing 
conditions, in the sense that we impose $\chi^{a}\!=\!0$ for all possible 
values of $\varphi$. This determines implicitly $m$ functions 
$\theta^{a}\!=\!\theta^{a}(\varphi)$ if we assume that the matrix
\beq
D^{a}\q_{b}
\~:=\~(\chi^{a}\papar{S^{i}})\~
(\Bar{\varphi}^{0i}\papar{\theta^{b}})\q_{\varphi}
\label{fpdet}
\eeq 
is invertible. (Note that unlike ordinary gauge--fixing, 
the BRST--invariants $S^{i}$ will depend on gauge-fixing conditions
$\chi^{a}\!=\!0$ by construction.) Then
\beq
0\=(\chi^{a}\papar{\varphi^{j}})
\=(\chi^{a}\papar{S^{i}})(S^{i}\papar{\varphi^{j}})
\~\equi{\e{sdiffwrtphi}+\e{fpdet}}\~
(\chi^{a}\papar{S^{i}})\~\Sigma^{0i}\q_{j}+D^{a}\q_{b}
\~(\theta^{b}\papar{\varphi^{j}}) \~. \label{chidiffwrtphi}
\eeq
Now we can use \eq{chidiffwrtphi} to rewrite \eq{sdiffwrtphi} as
\beq
(S^{i}\papar{\varphi^{k}})
\= P^{i}\q_{j}\~\Sigma^{0j}\q_{k}\~, 
\label{nextthing01}
\eeq
where we have defined the idempotent
\beq
Q^{i}\q_{j}\~:=\~
(\Bar{\varphi}^{0i}\papar{\theta^{a}})\q_{\varphi}\~
(D^{-1})^{a}\q_{b}\~
(\chi^{b}\papar{S^{j}})
\~,\qquad\qquad Q\=Q^{2}\~,\label{q01}
\eeq
and its complementary idempotent
\beq
P\~:=\~1-Q\=P^{2}\~,\qquad\qquad PQ\=0\=QP\~.
\eeq
This in turn implies
\beq
 P^{i}\q_{j}\~(\Bar{\varphi}^{0j}\papar{\theta^{a}})\q_{\varphi}\=0\~,
\label{nextthing02}
\eeq
and
\beq
(\chi^{a}\papar{S^{i}})\~P^{i}\q_{j}\=0\~. \label{nextthing03}
\eeq
It follows that $S^{i}$ is gauge--invariant,
\beq
(S^{i}\papar{\varphi^{k}})\~R^{0k}\q_{a}(\varphi) 
\~\equi{\e{nextthing01}}\~
P^{i}\q_{j}\~\Sigma^{0j}\q_{k}\~R^{0k}\q_{a}(\varphi) 
\~\equi{\e{bevisinvlieeq01}}\~ 
P^{i}\q_{j}\~(\Bar{\varphi}^{0j}\papar{\theta^{b}})\q_{\varphi}\~
\Tilde{\mu}^{0b}\q_{a} 
\~\equi{\e{nextthing02}}\~0\~.\label{lastbrst01}
\eeq
All together, we have solved the $J$-dependent Noether identities 
\e{noetherid01} in the original field sector \wthot inverse Lie 
\eq{bevisinvlieeq01}. It is easy to check that the other conditions in 
the $J$-dependent classical master \eq{sainvol} are satisfied as well.

\section{Quantum Corrections}
\label{secqcoor}

\noi
In this Section~\ref{secqcoor} we look for a solution to the quantum master 
\eq{qmej} with a truncated one-loop Ansatz of the form
\beq
W^{J}\=S^{J}+\Hi M^{J}\~.
\eeq
Besides the classical master \eq{cmej}, the quantum master \eq{qmej} becomes
\bea
 (M^{J},S^{J}) +\Delta\q_{\rho}S^{J}&=&0\~, \label{qmej1}  \\
\Hf (M^{J},M^{J}) +\Delta\q_{\rho}M^{J} +\nu\q_{\rho} &=&0\~. \label{qmej2} 
\eea
We now assume for simplicity Darboux coordinates 
$\Gamma^{A}=\{\Phi^{\alpha},\Phi^{*}_{\alpha}\}$ with trivial density $\rho=1$
and trivial odd scalar $\nu\q_{\rho}=0$. We furthermore assume that the
one--loop contribution 
\beq
M^{J}\=M^{0}(\varphi) \label{ansatzsomemore}
\eeq
only depends on the original fields $\varphi^{i}$, and in particular, that the 
one--loop contribution is independent of all the external sources $J\p_{A}$ and
all the antifields $\Phi^{*}_{\alpha}$. Then \eq{qmej2} is automatically
satisfied. The \eq{qmej1} reads in the sector proportional to $c^{a}$ 
\beq
(M^{0}(\varphi) \papar{\varphi^{i}}) R^{0i}\q_{a}(\varphi) 
+A_{a}^{0}(\varphi)\~\equi{\e{qmej2}}\~ 0 \~,\label{qmej2a} 
\eeq
where the formal anomaly function $A_{a}^{0}$ is defined in \eq{closedanomaly}.
Therefore
\bea
(M^{0}(\Bar{\varphi}^{0})\papar{\theta^{b}})\q_{\varphi}
&=&(M^{0}(\Bar{\varphi}^{0}) \papar{\Bar{\varphi}^{0i}})
\~(\Bar{\varphi}^{0i}\papar{\theta^{b}})\q_{\varphi}
\~\equi{\e{lieeq01}}\~
(M^{0}(\Bar{\varphi}^{0}) \papar{\Bar{\varphi}^{0i}})\~ 
R^{0i}\q_{a}(\Bar{\varphi}^{0})\~\lambda^{0a}\q_{b}(\theta,\varphi) \cr
&\equi{\e{qmej2a}}&
-A_{a}^{0}(\Bar{\varphi}^{0})\~\lambda^{0a}\q_{b}(\theta,\varphi)\~.
\label{qmej2b} 
\eea
Comparing with the differential \eq{epde01}, we conclude that a solution to 
the differential \eq{qmej2b} is 
\beq
M^{0}(\Bar{\varphi}^{0}) \= M^{0}(\varphi) -\ln E^{0}(\varphi,\theta)\~.
\eeq
The partition function reads
\beq
{\cal Z}^{\Psi}[J] 
\= \int \!  [d\Phi] \~\exp\left[M^{0}(\varphi)+\Ih 
S^{J}(\Phi,\Phi^{*}=\frac{\partial \Psi}{\partial\Phi}) \right]\~,
\label{zet02}
\eeq
where it is implicitly understood in \eq{zet02} that the field multiplet
\beq
\Phi^{\alpha}\=\{\varphi^{i};c^{a};\Bar{c}\q_{a};\pi\q_{a}\}
\eeq
now includes non-minimal fields for gauge-fixing purposes; namely a 
Faddeev--Popov antighost $\Bar{c}\q_{a}$ and a Nakanishi--Lautrup Lagrange
multiplier $\pi\q_{a}$; and it is furthermore implicitly understood that the 
minimal $S^{J}$ action \e{minsjaction} in \eq{zet02} has been replaced with 
the non-minimal action 
\beq
S^{J}\quad \longrightarrow\quad  S^{J}+ \Bar{c}^{*a}\pi\q_{a}\~.
\eeq
The partition function is independent of the gauge fermion $\Psi=\Psi(\Phi)$,
where the Faddeev-Popov matrix
\beq
\Delta^{a}\q_{b}\~:=\~
(\papal{\Bar{c}\q_{a}} \Psi \papar{\varphi^{i}})\~R^{0i}\q_{b}
\eeq 
is invertible; and where $\eps(\Psi)=1$ and  $\gh(\Psi)=-1$.

\section{Orbit Method}
\label{secorb}

\noi
In this Section~\ref{secorb}, we introduce the gauge parameter $\theta^{a}$
into the antisympletic phase space. Let us consider irreducible (possibly open)
theories in the minimal sector of the antisymplectic phase space
\beq
\Gamma\p_{\min}\~:=\~\{ \varphi^{i},\varphi_{i}^{*}; c^{a},c_{a}^{*}\}\~.
\label{minsector}
\eeq
The action in the minimal sector is of the form
\beq
S^{0}_{\min}\=\So^{0}(\varphi)+ \varphi_{i}^{*} \~R^{0i}\q_{a}(\varphi) \~c^{a} 
+ \ldots \~. \label{smin01}
\eeq
We assume for simplicity from now on that the underlying groupoid structure 
is independent of the external sources $J\p_{A}$. The main new feature in this 
Section~\ref{secorb} is that the gauge parameters $\theta^{a}$ and their
antifields $\theta_{a}^{*}$ are included into the total antisymplectic phase
space as active participants
\beq
\Gamma\p_{\tot}\~:=\~\{  \Gamma\p_{\min};\theta^{a},\theta_{a}^{*}\}\~.
\eeq
The action in the total sector is of the form
\beq
S^{0}_{\tot}\=S^{0}_{\min} -\theta^{*}_{a}\~
\Tilde{\mu}^{0a}\q_{b}(\theta,\varphi)\~c^{b} 
+\ldots\~. \label{stot01}
\eeq
The $S^{i}$ functions are of the form
\beq
S^{i}\= \Bar{\varphi}^{0i}(\varphi,\theta)
+\varphi_{j}^{*}\~K^{ij}\q_{a}(\varphi,\theta)\~c^{a}
-\theta_{a}^{*}\~K^{ia}\q_{b}(\varphi,\theta)\~c^{b}+ \ldots  \~. \label{si01}
\eeq
The set of classical master eqs.\ \e{sainvol} leads to a hierarchy of 
equations: 
(i) The Noether identity \e{noetherid01} in the sector proportional to $c^{a}$:
\beq
(\So^{0}\papar{\varphi^{i}})\~R^{0i}\q_{a}(\varphi) \=0\~.\label{noetherid02}
\eeq
(ii) An open version of inverse Lie \eq{bevisinvlieeq01} in the sector 
proportional to $J\p_{i}c^{a}$:
\beq
(\Bar{\varphi}^{0i}\papar{\varphi^{j}})\q_{\theta}\~ R^{0j}\q_{a}(\varphi) 
-(\Bar{\varphi}^{0i}\papar{\theta^{b}})\q_{\varphi}\~\Tilde{\mu}^{0b}\q_{a}
+  (\So^{0}\papar{\varphi^{j}})\~K^{ij}\q_{a}(-1)^{\eps_{i}}\=0\~.\label{jaycee01}
\eeq
Or equivalently, if one multiplies \eq{jaycee01} from left with the matrix 
$(\varphi^{i}\papar{\Bar{\varphi}^{0j}})\q_{\theta}$, one gets
\bea
R^{0i}\q_{a}(\varphi)
+(\varphi^{i}\papar{\theta^{b}})\q_{\Bar{\varphi}^{0}}\~\Tilde{\mu}^{0b}\q_{a}
&\equi{\e{jaycee01}}&
-(\varphi^{i}\papar{\Bar{\varphi}^{0k}})\q_{\theta}\~
(\So^{0}\papar{\varphi^{j}})\~K^{kj}\q_{a}(-1)^{\eps_{k}} \cr
&=&-(\So^{0}\papar{\varphi^{j}})
\~(\varphi^{i}\papar{\Bar{\varphi}^{k0}})\q_{\theta}\~K^{kj}\q_{a}
(-1)^{(\eps_{i}+\eps_{k})\eps_{j}+\eps_{k}}  \~. \label{jaycee02}
\eea
(iii) In the sector proportional to $J\p_{i}J\p_{j}c^{a}$, one gets
\bea
(\Bar{\varphi}^{0i}\papar{\varphi^{k}})\q_{\theta}\~\left( K^{jk}\q_{a} 
+(\varphi^{k}\papar{\theta^{b}})\q_{\Bar{\varphi}^{0}}\~K^{jb}\q_{a} \right)
&=& (\Bar{\varphi}^{0i}\papar{\varphi^{k}})\q_{\theta}\~K^{jk}\q_{a}
 -(\Bar{\varphi}^{0i}\papar{\theta^{b}})\q_{\varphi}\~K^{jb}\q_{a} \cr
&=&(-1)^{(\eps_{i}+1)(\eps_{j}+1)}(i\longleftrightarrow j)\~.
\label{jayjaycee01}
\eea
{}Firstly, note that the replacement of the closed inverse Lie
\eq{bevisinvlieeq01} with the open inverse Lie \eq{jaycee01} still allows for
essentially the same construction of the BRST--invariant $S^{i}$ from
Section~\ref{secbrstinv}. The only difference is that the off-shell
BRST--invariance \e{lastbrst01} turns into an on-shell BRST--invariance
\beq
(S^{i}\papar{\varphi^{k}})\~R^{0k}\q_{a}(\varphi) 
\~\equi{\e{nextthing01}+\e{jaycee01}+\e{nextthing02}}\~
-P^{i}\q_{j}\~(\So^{0}\papar{\varphi^{k}})\~K^{jk}\q_{a}(-1)^{\eps_{j}}\~.
\label{lastbrst02}
\eeq
Secondly, let us now consider a gauge orbit
\beq
\Bar{\varphi}^{0i}\=f^{0i}(\varphi,\theta)
\qquad\Longleftrightarrow\qquad 
 \varphi^{i}\=((f^{0})^{-1})^{i}(\Bar{\varphi}^{0},\theta)\~,
\eeq
and composed action
\beq
\Bar{\So}(\Bar{\varphi}^{0},\theta)
\~:=\~\So^{0}((f^{0})^{-1}(\Bar{\varphi}^{0},\theta))\~.
\label{composedaction01}
\eeq
Multiplying \eq{jaycee01} with
$(\Bar{\So}\papar{\Bar{\varphi}^{0i}})\q_{\theta}$ yields
\bea
0&\equi{\e{noetherid02}+\e{jaycee01}}&
(\Bar{\So}\papar{\theta^{b}})\q_{\Bar{\varphi}^{0}}\~\Tilde{\mu}^{0b}\q_{a}
+(\Bar{\So}\papar{\Bar{\varphi}^{i}})\q_{\theta}\~(\So^{0}\papar{\varphi^{j}})\~
K^{ij}\q_{a}(-1)^{\eps_{i}} \cr
&\equi{\e{jayjaycee01}}&
(\Bar{\So}\papar{\theta^{b}})\q_{\Bar{\varphi}^{0}}\~\Tilde{\mu}^{0b}\q_{a}
-(\Bar{\So}\papar{\Bar{\varphi}^{0i}})\q_{\theta}\~(\So^{0}\papar{\varphi^{j}})\~
(\varphi^{j}\papar{\theta^{b}})\q_{\Bar{\varphi}^{0}}\~
K^{ib}\q_{a} (-1)^{\eps_{i}} \cr
&=&(\Bar{\So}\papar{\theta^{b}})\q_{\Bar{\varphi}^{0}}\left(\Tilde{\mu}^{0b}\q_{a}
-(\Bar{\So}\papar{\Bar{\varphi}^{0i}})\q_{\theta}\~
K^{ib}\q_{a}(-1)^{\eps_{i}(\eps_{b}+1)} \right)\~.\label{helpinghand01}
\eea
Assuming that the $\Tilde{\mu}^{0b}\q_{a}$ matrix is an invertible matrix,
we deduce that the action \e{composedaction01} is gauge invariant
\beq
(\Bar{\So}\papar{\theta^{a}})\q_{\Bar{\varphi}^{0}}\~\equi{\e{helpinghand01}}\~
 0\~,\label{gaugeinv01}
\eeq
at least sufficiently close to the classical trajectories
$\So^{0}\papar{\varphi^{i}} \approx 0$.
We next introduce shifted structure functions
\beq
\Tilde{K}^{ij}\q_{a}
\~:=\~(\varphi^{i}\papar{\Bar{\varphi}^{0k}})\q_{\theta} \left( K^{kj}\q_{a} 
+(\varphi^{j}\papar{\theta^{b}})\q_{\Bar{\varphi}^{0}}\~K^{kb}\q_{a} \right)
(-1)^{(\eps_{i}+\eps_{k})\eps_{j}+\eps_{k}}  
\~\equi{\e{jayjaycee01}}\~
-(-1)^{\eps_{i}\eps_{j}}(i\longleftrightarrow j)\~.\label{kaytilde01}
\eeq
Eqs.\ \e{jaycee02}, \es{gaugeinv01}{kaytilde01} imply an 
$(i\leftrightarrow j)$ symmetric version of \eq{jaycee02}: 
\beq
R^{0i}\q_{a}(\varphi)
+(\varphi^{i}\papar{\theta^{b}})\q_{\Bar{\varphi}^{0}}\~\Tilde{\mu}^{0b}\q_{a}
\=-(\So^{0}\papar{\varphi^{j}})\~\Tilde{K}^{ij}\q_{a}\~.\label{symmarum01}
\eeq
We stress that \eq{jaycee02}, or equivalently \eq{symmarum01}, can be viewed
as an open version of the inverse Lie \eq{invlieeq01} for quasi--groups.

\section{Gauge--Invariants}
\label{secgi}

\noi
In this Section~\ref{secgi}, we construct on-shell gauge--invariants $\xi^{I}$.
See also Section 4.1 in \Ref{bb10}. Let $\chi^{a}\!=\!\chi^{a}(\Bar{\varphi}^{0})$
be $m$ gauge-fixing conditions, in the sense that we impose $\chi^{a}\!=\!0$
for all possible values of $\Bar{\varphi}^{0i}$. The gauge-fixing conditions
leave $n-m$ gauge--invariants $\xi^{I}$ unconstrained:
\beq
\chi^{a} (\Bar{\varphi}^{0})\= 0 \qquad\Longleftrightarrow\qquad 
\Bar{\varphi}^{0i}\=g^{i}(\xi)\~. \label{gijoe01}
\eeq
(Again, note that unlike ordinary gauge--fixing, the gauge--invariants 
$\xi^{I}$ will depend on gauge-fixing conditions $\chi^{a}\!=\!0$ by 
construction.) Recalling that
\beq
\Bar{\varphi}^{0i}\=f^{0i}(\varphi,\theta)\~,
\eeq
we can now reparametrize the original variable $\varphi^i$ as
\beq
(\xi^{I},\theta^{a})
\qquad \longrightarrow\qquad
\varphi^{i} \= ((f^{0})^{-1})^{i}(g(\xi),\theta) \~.\label{mappemap01}
\eeq
In words, the coordinates $(\xi^{I},\theta^{a})$ represent the decomposition 
of the $n$ original fields $\varphi$ in $n-m$ physical gauge--invariants 
$\xi^{I}$ and $m$ gauge variables $\theta^{a}$. The following rank conditions 
are assumed:
\beq
\rank(\Bar{\varphi}^{0i}\papar{\xi^{I}})
\=n-m\~, \qquad 
\rank(\Bar{\varphi}^{0i}\papar{\theta^{a}})\q_{\varphi}\=m \~. 
\eeq
We assume that there exists an inverse map to the reparametrization 
\e{mappemap01}
\beq
\varphi^{i} \qquad \longrightarrow\qquad \xi^{I}\=\xi^{I}(\varphi)\~,\qquad 
\theta^{a}\=\theta^{a}(\varphi)\~.
\eeq
The fact that $\xi^{I}$ are independent of $\theta^{a}$ is encoded via 
the relations
\beq
0\=(\xi^{I}\papar{\varphi^{i}})  
\~(\varphi^{i}\papar{\theta^{b}})\q_{\Bar{\varphi}^{0}}\~. \label{rydberg}
\eeq
Next, let us consider a source-dependent master action of the form
\beq
S^{J}\=S^{0}_{\min}+ J\p_{I}\~ \Xi^{I}
\eeq
in the minimal sector \e{minsector}. Here the BRST--invariants 
\beq
\Xi^{I}\= \xi^{I} + \varphi_{i}^{*}\~ \Bar{K}^{Ii}\q_{a} \~c^{a}+\ldots
\label{xibrstinv}
\eeq
are deformations of the gauge invariants $\xi^{I}$.
The set of classical master eqs.\ \e{sainvol} in the minimal sector reads 
\beq
(S^{0}_{\min},S^{0}_{\min})\=0\~,\qquad(S^{0}_{\min},\Xi^{I})\=0\~,\qquad 
(\Xi^{I},\Xi^{L})\=0\~. \label{sainvolmin}
\eeq
The second and third involution eqs.\ \e{sainvolmin} imply, among other things, 
that
\beq
(\xi^{I}\papar{\varphi^{i}})\~R^{0i}\q_{a}(\varphi) + 
(\So^{0}\papar{\varphi^{i}})\~\Bar{K}^{Ii}\q_{a}(-1)^{\eps_{I}}
\=0\~, \label{symmarum03}
\eeq
and
\beq
(\xi^{I}\papar{\varphi^{i}})\~\Bar{K}^{Li}\q_{a} 
\=(-1)^{(\eps_{I}+1)(\eps_{L}+1)}(I\longleftrightarrow L)\~,\label{kaytilde02}
\eeq
respectively.

\noi
Moreover, we can also derive \eqs{symmarum03}{kaytilde02} from the 
orbit method of Section~\ref{secorb}. If we use the open version of the 
inverse Lie \eq{symmarum01}, we get
\beq
(\xi^{I}\papar{\varphi^{i}}) \left(R^{0i}\q_{a}(\varphi) + 
(\So^{0}\papar{\varphi^{j}})\~\Tilde{K}^{ij}\q_{a}\right)
\~\equi{\e{symmarum01}+\e{rydberg}}\~
 0\~.\label{symmarum02}
\eeq
Eq.\ \e{symmarum02} shows that $\xi^{I}$ are gauge--invariant on--shell. If we
now identify
\beq
\Bar{K}^{Ii}\q_{a}\~=\~-(\xi^{I}\papar{\varphi^{j}})\~
\Tilde{K}^{ij}\q_{a} (-1)^{\eps_{I}(\eps_{i}+1)}
\~\equi{\e{kaytilde01}+\e{rydberg}}\~
(\varphi^{i}\papar{\Bar{\varphi}^{0k}})\q_{\theta}\~
(\xi^{I}\papar{\varphi^{j}})\~
K^{kj}\q_{a} (-1)^{(\eps_{I}+1)(\eps_{k}+1)}
\~,\label{kaybar}
\eeq
then \eqs{symmarum02}{kaytilde01} become the classical master 
\eqs{symmarum03}{kaytilde02}, respectively.

\noi
{}Finally, let us use the $g^{i}$ functions from \eq{gijoe01} and the
BRST--invariants \e{xibrstinv} to define a new set of BRST--invariants 
\beq
S^{i} \=g^{i}(\Xi)\~\equi{\e{xibrstinv}}\~g^{i}(\xi) 
+ (g^{i}(\xi)\papar{\xi^{I}})\~\varphi_{j}^{*}\~\Bar{K}^{Ij}\q_{a} \~c^{a}+\ldots\~,
\label{gijoe02}
\eeq
which we pair with the minimal action $S^{0}_{\min}$. It follows that
$\{S^{0}_{\min};S^{i}\}$ satisfies the set of classical master eqs.\ \e{sainvol}, 
because $\{S^{0}_{\min};\Xi^{I}\}$ does, \cf \eq{sainvolmin}. This means that 
$S^{i}$ is a minimal analogue to the $S^i$ function \e{si01} without 
the $\{\theta^{a};\theta^{*}_{a}\}$ dependence (so that, \eg the $K^{ia}\q_{b}$  
structure functions are absent)
\beq
S^{i}\=\Bar{\varphi}^{0i}(\varphi)
+\varphi_{j}^{*}\~K^{ij}\q_{a}(\varphi)\~c^{a}+ \ldots  \~. \label{si01min}
\eeq
Here $\Bar{\varphi}^{0i}=g^{i}(\xi)$ and  
\beq
K^{ij}\q_{a}
\=(g^{i}(\xi)\papar{\xi^{I}})\~\Bar{K}^{Ij}\q_{a}
(-1)^{(\eps_{I}+\eps_{i})(\eps_{j}+1)}\~.
\eeq
Moreover $S^i$ and $K^{ij}\q_{a}$ satisfy minimal versions of the corresponding 
formulas from Section~\ref{secorb}.

\vspace{0.8cm}

\noi
{\sc Acknowledgement:}~K.B.\ would like to thank K.P.~Zybin and the Lebedev
Physics Institute for warm hospitality. The work of I.A.B.\ is supported by
grants RFBR 11--01--00830 and RFBR 11--02--00685. The work of K.B.\ is
supported by the Grant Agency of the Czech Republic (GACR) under the grant 
P201/12/G028.

\end{document}